\documentclass[twocolumn, showpacs, amsmath,amsfonts,prb]{revtex4-1}

\usepackage{graphicx}
\usepackage{amsmath}
  \usepackage{amssymb}
  \usepackage{bm}
  
\usepackage[cp1251]{inputenc}
\usepackage[T1,T2A]{fontenc}

\usepackage{color}

\begin{document}
\date{\today}
  \title{Resonant Wood's anomaly diffraction condition \\in 
  dielectric and
  plasmonic grating structures}

   \author{M. M. Voronov}
   \email{mikle.voronov@coherent.ioffe.ru}
    \affiliation{Ioffe Institute, St.Petersburg, Russia}

\begin{abstract}
The general features of the light scattering resulting in the
so-called resonant Wood's anomalies in the reflection and
transmission spectra are described using the effective parameters
of a quasi-guided mode. The expression determining the
spectral angular dependence of Wood's anomaly in the case of
plasmonic grating structures is given and compared to 
the analogous expression for dielectric grating structures. 
Comparison of the resonant Wood's anomalies (RWA) with the 
surface plasmon-polariton resonances (SPPRs) is discussed as well.
\end{abstract}

 \maketitle
 
\section*{Introduction}
In the recent years in condensed-matter physics much attention has
been focused on the study of the strong optical effects arising at
the exterior surfaces of microstructures, in particular, of the
optical resonances, which are due to the periodicity of a grating
structure and can be essentially modified by intrinsic material
excitations, e.g., such as plasmons [\onlinecite{Raether,Maier}]. 
As a result, these resonances lead to a high intensity of scattered
light around a resonant frequency, which in some cases can be
approximately determined by the corresponding diffraction condition. 
The well-known examples of surface optical resonances are the 
resonant Wood's anomaly [\onlinecite{Hessel,Sarrazin,Maystre}] and
surface plasmon-polariton resonance [\onlinecite{Ghaemi,Wedge,Gao,Sha}],
that manifest themselves in the optical spectra as sharp peaks 
whose position and height depend on the spatial configuration of the
structure and the dielectric function of the constituent materials.
The origination of RWA is associated with the formation of 
a quasi-guided mode in the near-surface layer [\onlinecite{Fano41}].

Usually, in the case of two- and three-dimensional periodic structures 
it is impossible to derive explicit analytical solutions suitable for
analysis and one has to resort to methods based on using effective
parameters. The approach using the effective parameters of a 
quasi-guided mode, which allows one to formulate the diffraction 
condition for RWA appearing in optical spectra of dielectric grating
structures (DGSs), is shortly described in [\onlinecite{Voronov_PRB2014}].
In this diffraction condition a single scattering vector is taken into
account explicitly, while the rest of the scattering can be included 
in the effective parameters of a quasi-guided mode. In the present
paper we generalize this theory to the grating structures containing 
metallic components with permittivity described by the Drude-Lorentz 
model and take into account the frequency dispersion of the transverse 
component of the wave vector in a waveguide layer.

Plasmonic grating structures (PGSs) in the simplest case are two- or
three-layer structures with a periodic arrangement of metal
components in the outside layer (I), or with a metallic layer from a
side of the substrate (II), or combining the both types together
(III). In this paper we consider PGS of the types I and III, in which
case we give the expression for the positions of Wood's anomaly
peaks and the corresponding dispersion relation.

\section{General diffraction condition}
Using the theory developed in [\onlinecite{Voronov_PRB2014}], 
we write the diffraction condition for RWA as ${\bf k}+{\bf G}={\bf
k'}$, where ${\bf G}$ is a reciprocal lattice vector and ${\bf k'}$
is the wave vector of the wave propagating in a surface layer of the
structure and corresponding to a quasi-guided mode
(\textquotedblleft quasi\textquotedblright\: because of wave 
scattering to the outside). The magnitude of 
$\bf{k'}$ is $k'=|{\bf k'}|=\omega\sqrt{\tilde{\varepsilon}}/c$, 
where $\omega$ is the light frequency, and $\tilde{\varepsilon}$ 
is the effective dielectric function of the surface layer. The wave
vector ${\bf k}$ can be represented as ${\bf k} = {\bf k_{||}}+{\bf k}_z$, 
where $k_{||}=k_0\sin\theta$, $k_0=\omega/c$. By combining these
equations one gets
\begin{equation}
 {k_0}^2(\tilde{\varepsilon}-{\sin}^{2}\theta)-2k_{0}G\sin{\theta}
\cos{\varphi}-G^2-{k_z}^2=0\:,
\end{equation}
where $\varphi$ is the angle between vectors ${\bf k_{||}}$ and
${\bf G}$, and $\theta$ is the outgoing angle of the light with
respect to the normal to the surface. For a light wave propagating
in the surface layer the transverse component $k_z$ can be taken
constant, as it occurs in typical waveguides. From Eq. (1) one gets
\begin{equation}
\sin\theta=\frac{G}{k_0}\left(\pm\sqrt{\frac{k_0^2\tilde{\varepsilon}-k_z^2}{G^2}-\sin^2\varphi}-\cos\varphi\right)
\end{equation}
It must satisfy the inequality $0<\sin\theta<1$, while the
expression under the square root sign be nonnegative. Formally, 
there are two solutions for $\sin\theta$ and, as the analysis shows,
there are two cases: i) one solution of Eq. (2), with the plus
sign before the square root, if
$-k_0(k_0+2G\cos\varphi)<k_z^2+G^2-k_0^2\tilde{\varepsilon}<0$
(which is possible for any sign of $\cos\varphi$), or if
$k_0^2\tilde{\varepsilon}=k_z^2+G^2$, then $-k_0<2G\cos\varphi<0$;
and ii) two solutions of Eq. (2) (both signs before the radical),
if $\sqrt{k_z^2+G^2-k_0^2\tilde{\varepsilon}}\leq G|\cos\varphi|$,
$k_z^2+G^2>k_0^2\tilde{\varepsilon}$, $\cos\varphi <0$.

The diffraction condition for RWA corresponding to the same 
quasi-guided mode but to different scattering vectors,
${\bf{G}}_1\neq {\bf{G}}_2$, satisfy the equations
\begin{equation}
|{\bf k}_1+{\bf G}_1|=|{\bf k}_2+{\bf
G}_2|=\frac{\omega}{c}\sqrt{\tilde{\varepsilon}}\:,\:\:
|k_{1z}|=|k_{2z}|\:,
\end{equation}
and, consequently, two Eqs. (1) with the same values of 
$\tilde{\varepsilon}$ and $|k_z|$. However, Eqs. (1) and  (3) define
not only the condition for escape of the light wave from the grating 
structure at the angle $\theta$, but also may give the condition at 
which the incident (at the angle $\theta_0$) light wave is scattered 
to produce a quasi-guided mode. In particular, Eq. (3) is fulfilled 
in the case of the same scattering vector, 
${\bf G}_1={\bf G}_2={\bf G}$ (i.e. $G_1=G_2$,
$\cos\varphi_1=\cos\varphi_2$). 
From Eqs. (2) and (3) for incident and reflected waves we infer that
the resonant Wood's anomaly may arise at the diffraction angle 
$\theta=\theta_0$, that is realized in experiment 
[\onlinecite{Voronov_PRB2014}] and numerical calculations 
[\onlinecite{Tikhodeev2002}].
(The case $\theta\neq\theta_0$ satisfying either item ii) 
or Eq. (3) at ${\bf G}_1\neq{\bf G}_2$ can hardly be realized, 
because the intensity of the corresponding wave should be much
smaller than that of the specularly scattered wave.)

\section{Spectral angular dependencies and dispersion relation}

Let us turn to the study of the resonant Wood's anomalies in the
reflection and transmission spectra for PGSs. We write the dielectric
function of the metal components of the structure, e.g., of
parallel metal strips on the surface of a dielectric plate as
$$\varepsilon(\omega)=1-\frac{\omega_p^2}{\omega^2+i\gamma\omega}+\chi_1(\omega)+i\chi_2(\omega)\:.$$
Here $\omega_p$ is the plasma frequency with which the metal
components are characterized, $\gamma$ is the damping constant, and
$\chi=\chi_1+i\chi_2$ is a contribution in susceptibility due to
interband electronic transitions [\onlinecite{Pinchuk}]. In the 
simplest approximation one can take $\chi_1$ to be constant and 
neglect the imaginary part $\chi_2$ (far from the resonance frequency
of the metal). It is also possible to neglect $\gamma$, that can be
partly justified by the fact that incident radiation supports a 
quasi-guided mode and therefore to some extent compensates for the 
energy losses caused by absorption. Thus, the effective dielectric 
function of the waveguide surface layer has the form
$\tilde{\varepsilon}=\varepsilon_0-(\omega_p/\omega)^2$, where
$\varepsilon_0$ takes into account the dielectric material of the
grating structure as well. By substituting $\tilde{\varepsilon}$
into Eq. (1), we get the expression for the Wood's anomaly spectral
peak position
\begin{equation}
\lambda_W=\frac{2\pi}{G}\frac{{\sqrt{(\sin{\theta}\cos{\varphi})^{2}+(b
-\sin^{2}{\theta})(1+a)}}-\sin\theta\cos\varphi}{1+a}\:,
\end{equation}
where $a=(k_z/G)^2+\omega_p^2/(Gc)^2$ and $b=\varepsilon_0$ ($b>1$).

Equation(4) has the same form as the analogous expression in the
case of DGSs [\onlinecite{Voronov_PRB2014}], but where the parameter
$a$ defines a different quantity. If to set $\omega_p=0$, the dependence
of $\tilde{\varepsilon}$ on the frequency $\omega$ vanishes, hence
$\tilde{\varepsilon}=n_{eff}^2$ (where $n_{eff}$ is the effective
refractive index independent of $\omega$) and $a=(k_z/G)^2$ as it is
in the case of a purely dielectric structure. On the whole, Eq. (4) 
for $\lambda_W(\theta)$ at different values of $a$ and $b$ is an 
almost linear function of $\theta$ with a small curvature and with 
the angle of slope, $\partial\lambda_W/\partial\theta$, depending on
a relation between the values of $a$, $b$ and $\cos{\varphi}$.

For a light wave propagating in the surface layer the transverse 
component $k_z$ can be taken approximately constant, as it occurs in 
rectangular dielectric waveguides, however, as the surface layer is 
not spatially homogeneous, one should take into account the frequency
dispersion of the $k_z$ component. Since in the case of PGS a correction
to the dielectric function (associated with the value of $\varepsilon_0$) 
is taken to be constant (in zero order in $\omega$), in a decomposition of
$k_z^2$ we restrict ourselves to the quadratic order in $k_0$, that is 
$k_z^2=\alpha +\beta k_0 + \gamma k_0^2$ (which in particular
corresponds to the linear approximation $|k_z|=|k_z^{(0)}|+k_0\delta$), 
where $\alpha$, $\beta$, $\gamma$, $\delta$ are some constants. 
Substituting the above expression for $k_z^2$ into Eq. (1), we obtain 
\begin{equation}
{k_0}^2(\epsilon-{\sin}^{2}\theta)-k_{0}(\beta+2G\sin{\theta}
\cos{\varphi})-C=0\:, 
\end{equation}
where $\epsilon=\varepsilon_0-\gamma$ and $C=(\omega_p/c)^2+G^2+\alpha$
(so that, $\epsilon>1$, and $C>0$). Hence, after simple transformations,
one gets the following expression:
\begin{eqnarray}
\lambda_W & = & \frac{\pi}{C}\sqrt{(\beta+2G\sin{\theta}\cos{\varphi})^2 + 4(\epsilon
-\sin^{2}{\theta})C} \nonumber\\&&
-\frac{\pi}{C}(\beta+2G\sin{\theta}\cos{\varphi})\:, 
\end{eqnarray}
This expression is a generalization of Eq. (4), which also gives an almost
linear spectral angular dependence $\lambda_W(\theta)$. In particular, such
dependencies were experimentally obtained for hybrid opaline photonic crystals 
[\onlinecite{Voronov_PRB2014}]; these are well described by Eq. (4) derived
from Eq. (1) in which the effective dielectric function, $\tilde{\varepsilon}$, 
and the transverse wave-vector component, $k_z$, are set to constant values.
It may signify that the frequency dispersion of these quantities for such
structures (and possibly for other dielectric structures) is rather small.  
Since PGSs are characterized with a larger number of parameters and commonly
yield richer spectra than analogous dielectric structures, the former look
more likely to give the spectral angular dependencies different from those 
described by Eq. (4), and Eq. (6) should be used instead.

It is convenient to introduce the following notation:
$\eta_0 = G_x(\sqrt{\beta^2+4\epsilon(C-G_x^2)}-\beta)/[2(G_x^2-C)]$, where
$G_x=G\cos\varphi$. From the analysis of Eq. (6) one can obtain the following 
conditions for the cases of increasing and decreasing $\lambda_W$ with an 
increase in $\theta$:

a) $\partial\lambda_W/\partial\theta>0$ if simultaneously
$\cos\varphi<0$ and\\ $0<\sin\theta<\eta_0$;

b) $\partial\lambda_W/\partial\theta<0$ if simultaneously $\cos\varphi<0$ and\\
$\eta_0<\sin\theta<1$, and also if $\cos\varphi>0$;

c) $\partial\lambda_W/\partial\theta=0$ if simultaneously
$\cos\varphi<0$ and\\ $\theta=\arcsin\eta_0$.

A considerable simplification of the expression for $\lambda_W$ suitable for
making rough estimates in the case of 1D geometry of scattering ($|\cos\varphi|=1$)  
can be made by setting in Eq. (6) $\alpha=\beta=0$ ($|k_z|=\sqrt{\gamma}k_0$).
Hence, one gets $\lambda_W(\theta)=(2\pi/G)(\tilde{n}\pm\sin\theta)$, where 
$\tilde{n}=\sqrt{\tilde{\varepsilon}-\gamma}$ in the case of a DGS and
$\tilde{n}=\sqrt{\varepsilon_0-\gamma}$ for a PGS; the plus sign corresponds to
$\varphi = 180^0$, and the minus sign to $\varphi = 0$. 

Usually instead of the function $\lambda_W(\theta)$ one numerically calculates
the in-plane dispersion relation $\omega(k_{||})$, which in the framework of  
this approach can be obtained by solving Eq. (5) for $k_0$ :
\begin{equation}
k_0(k_{||})=\frac{\beta + \sqrt{\beta^2+4(\varepsilon_0-\gamma)(g(k_{||})+p+\alpha)}}{2(\varepsilon_0-\gamma)}\:,
\end{equation}
where $p=(\omega_p/c)^2$ and $g(k_{||})=k_{||}^2+2k_{||}G\cos\varphi +G^2$.
\begin{figure}[b]
\begin{center}
 \includegraphics[width=0.55\textwidth]{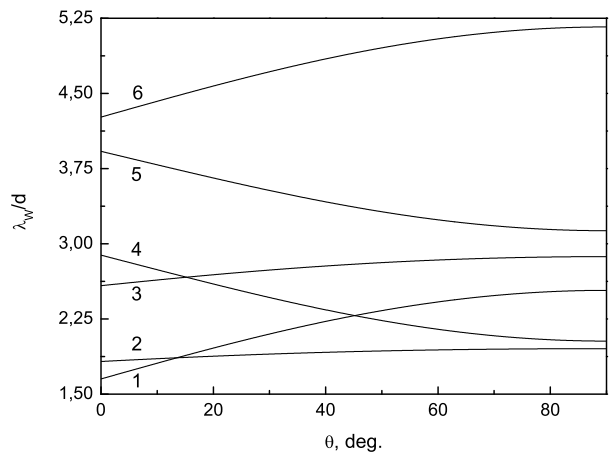}
 \end{center}
\caption{The spectral angular dependencies calculated by Eq. (6)
for different quasi-guided modes of the Wood's anomalies, with the
following parameters ($l=1$, $\beta=0$):  
$a_1=0.1$, $b_1=3$, $\varphi_1=\pi$; $a_2=5$, $b_2=20$, $\varphi_2=\pi$;
$a_3=2$, $b_3=20$, $\varphi_3=\pi$; $a_4=0.2$, $b_4=10$, $\varphi_4=0$;
$a_5=0.3$, $b_5=20$, $\varphi_5=0$; $a_6=0.1$, $b_6=20$, $\varphi_6=\pi$.
}
\end{figure}
\begin{figure}[t]
\begin{center}
 \includegraphics[width=0.55\textwidth]{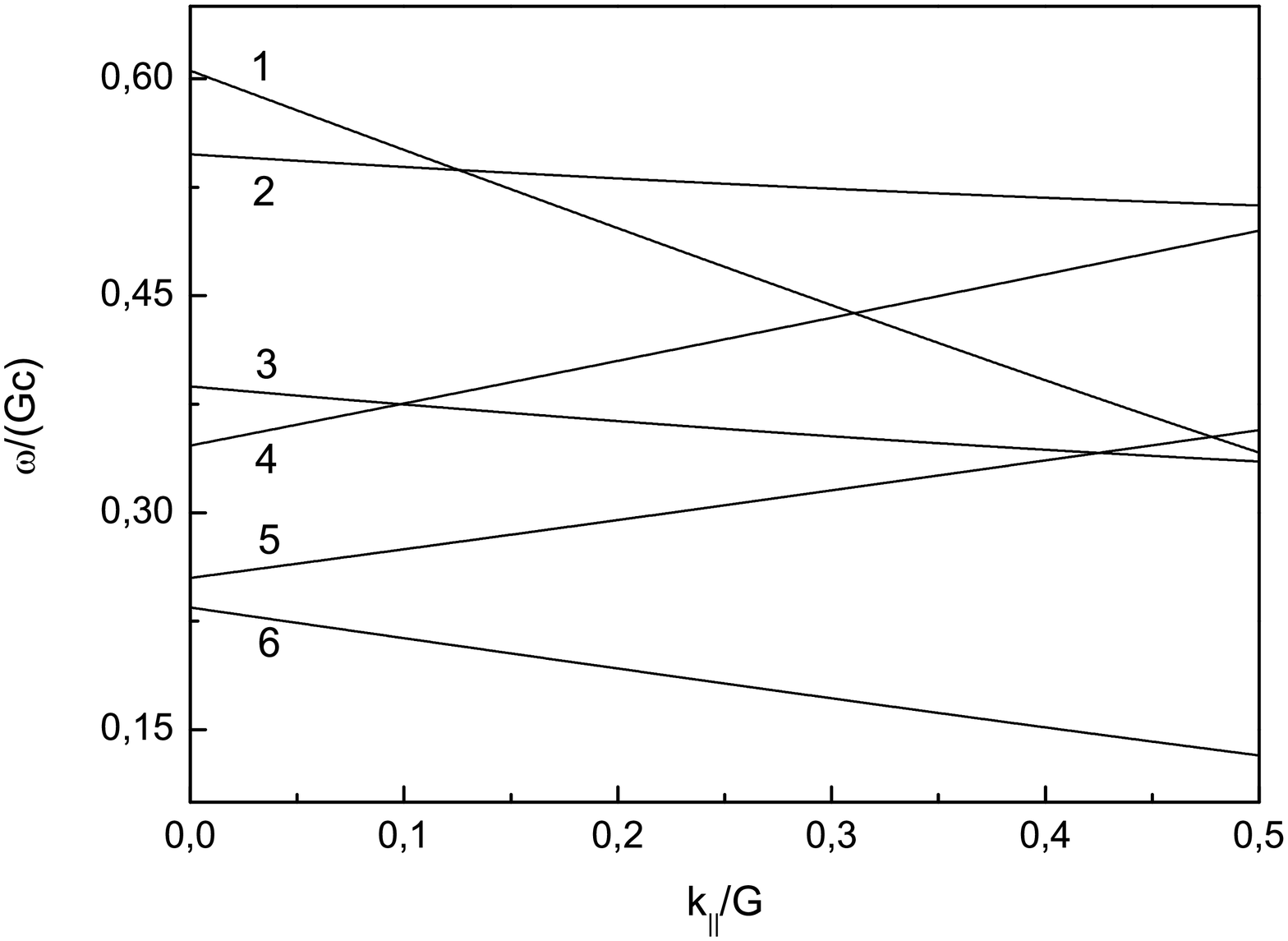}
 \end{center}
\caption{Dispersion curves plotted by using Eq. (7) for the same 
parameters as in Fig. 1.}
\end{figure}

Here we will illustrate the application of Eqs. (6) and (7) to several
sets of parameters $a=(\alpha +p)/G^2$ and $b=\varepsilon_0-\gamma$ 
used to describe the spectral angular dependencies of RWA in optical 
spectra of different grating structures periodic in one dimension. 
Figure 1 shows these dependencies in the zero diffraction order, 
$\theta=\theta_0$, when the scattering vector ${\bf G}$ ($G=2\pi l/d, l=1$)
lies in the incidence plane ($\varphi=0$ or $\varphi=\pi$), and Fig. 2 shows
the corresponding dispersion curves $\omega(k_{||})$. As is seen from Fig.~1, 
the Wood's anomalies can be apparent in a rather wide wavelength range 
limited to $\lambda_{max}=d(1+\sqrt{\varepsilon_m})$, where $d$ is the period
of the structure and $\varepsilon_m$ is the maximum value of the dielectric
constant. An important feature of RWA is the sign of the derivative 
$\partial\lambda_W/\partial\theta$. In many experimental spectra
$\partial\lambda_W/\partial\theta>0$, which is possible only in the case of the
backward scattering of light ($\cos\varphi<0$) for the diffraction angles 
$\theta<\arcsin\sqrt{b/a}$. In dielectric grating structures
the parameter $a=(k_z/G)^2\propto (nd/\lambda)^2$ is typically of the order 
of 1 or less, while the parameter $b=n^2_{eff}$ is greater than 2,
therefore the last inequality is satisfied for the whole range of angles $\theta$.
Typical spectral angular dependencies, $\lambda_W(\theta)$, observed in optical
spectra [\onlinecite{Voronov_PRB2014}] are some increasing concave functions of
$\theta$, that is 
$\partial\lambda_W/\partial\theta>0\:,\:\:\partial^2\lambda_W/\partial\theta^2<0\:,$ 
see also curves 1-3 and 6 in Fig. 1. Actually, as follows from the analysis
of Eq.~(4), in the case of backscattering of light at the condition $b>a$ 
the second derivative of $\lambda_W(\theta)$ remains negative up to quite 
large values of $\theta$.

Equation (4) allows one to determine the peak positions of RWA corresponding 
to a quasi-guided mode characterized by only two parameters, $a$ and $b$,
therefore we now use it instead of Eq. (6).
The peaks corresponding to different modes can greatly differ in intensity so that
in the frequency range of interest a single quasi-guided mode can dominate over 
the other ones. At the same time, one can expect a series of lines of the RWA
peaks, which give almost parallel curves $\lambda_W(\theta)$. This situation 
can be realized if for any angle $\theta$ the derivative 
$\partial\lambda_W/\partial\theta\approx const$ 
at a constant value of the parameter $a$, but at different values of $b$.
In the simplest case, where $|\cos\varphi|=1$, this condition is satisfied if 
$b(1+a)/(a\sin^2\theta)-1\gg a$; it is satisfactorily fulfilled
in a wide range of the angles $\theta$ and parameters $a$ and $b$.
As an example, Fig. 3 shows two sets of spectral angular dependencies plotted for
two different values of the $a$ parameter ($a=0.1$ and 1), while the $b$ parameter
varies in a wide range (from 5 to 20). 
\begin{figure}[h]
\begin{center}
 \includegraphics[width=0.53\textwidth]{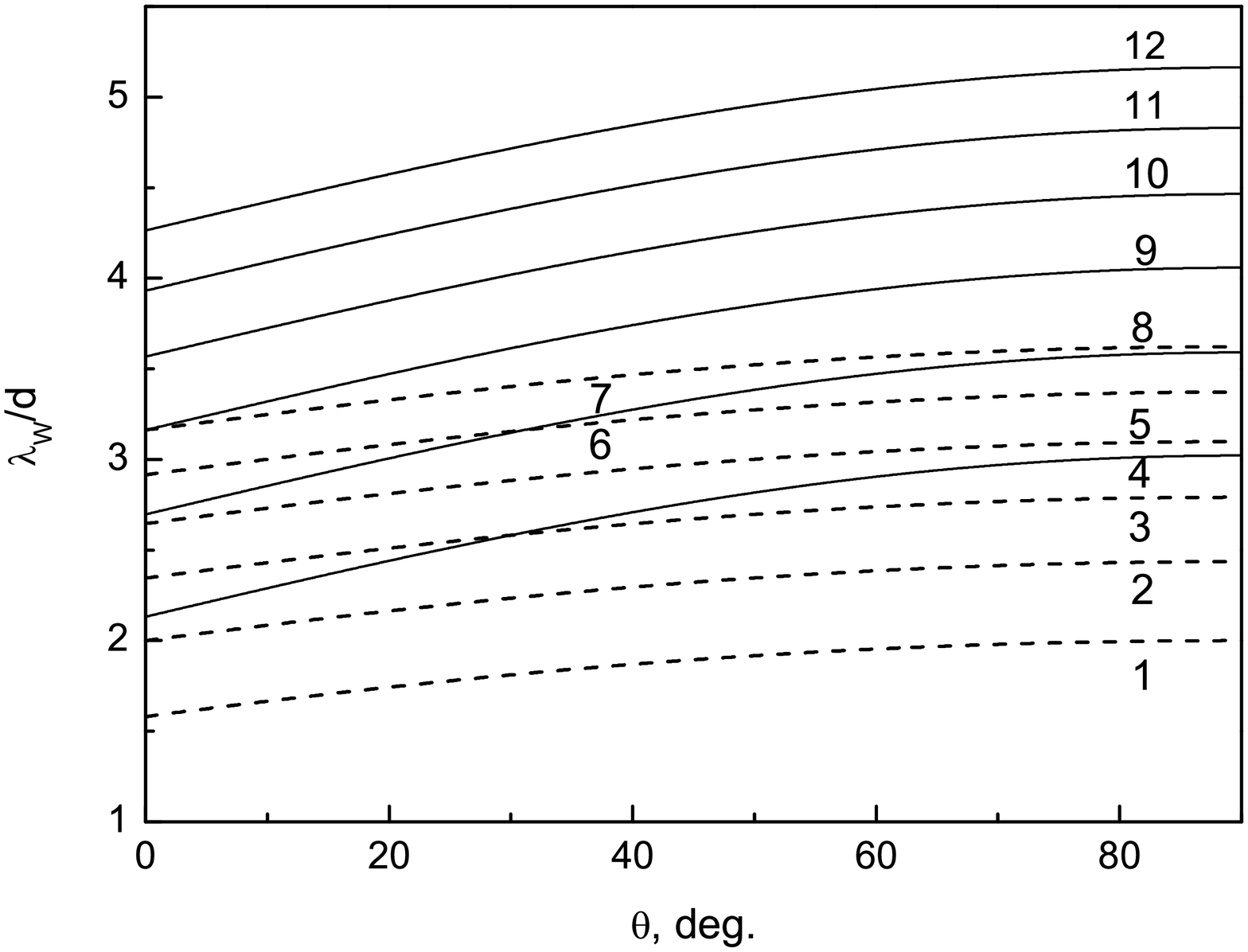}
 \end{center}
\caption{The spectral angular dependencies calculated by Eq. (4) for
different quasi-guided modes of the Wood's anomalies, with the
following parameters: $a_1=0.1$, $a_2=1$, $b_1=5$,
$b_2=8$, $b_3=11$, $b_4=14$, $b_5=17$, $b_6=20$,
$\varphi=\pi$. The sets of the
parameters under study are as follows: 1-($a_2$,$b_1$),
2-($a_2$,$b_2$), 3-($a_2$,$b_3$), 4-($a_1$,$b_1$), 5-($a_2$,$b_4$),
6-($a_2$,$b_5$), 7-($a_1$,$b_2$), 8-($a_2$,$b_6$),
9-($a_1$,$b_3$), 10-($a_1$,$b_4$), 11-($a_1$,$b_5$), 12-($a_1$,$b_6$).
}
\end{figure}

\section{Comparison to the surface plasmon-polariton resonances}

The resonant Wood's anomalies in PGSs should be distinguished from
the surface plasmon-polariton resonances, appearing in the 
reflection spectra provided that one (the A layer) of the layers
of the structure in a given frequency range has a negative 
dielectric constant, $\varepsilon_1<0$ (for a metal layer) 
and $|\varepsilon_1|>\varepsilon_2>0$, where $\varepsilon_1$ and
$\varepsilon_2$ are the dielectric constants of the $A$ and $B$ layers,
respectively. The in-plane ($XY$) component of the wave vector of
the surface wave is written as $q_{||}=k_0\sqrt{\varepsilon_s}$, where
$\varepsilon_s=\varepsilon_1\varepsilon_2/(\varepsilon_1+\varepsilon_2)$
and satisfies the diffraction condition ${\bf q_{||}} = {\bf k_{||}}+{\bf G}$
[\onlinecite{Ghaemi}].
The transverse component of the wave vector of the surface wave is
given by $q_{z,j}=(k_0^2\varepsilon_j-{q_{||}}^2)^{1/2}$, where
$j=1,2$, hence, taking into account the equality ${\bf q_{||}}={\bf
k_{||}}+{\bf G}$, after some transformations  one gets two equations
similar to Eq. (1), namely
$$
{k_0}^2(\varepsilon_j-{\sin}^{2}\theta)-2k_{0}G\sin{\theta}
\cos{\varphi}-G^2-{q_{z,j}}^2=0\:,
$$
where $\varepsilon_j$ is $\varepsilon_1=1-(\omega_p/\omega)^2$ and
$\varepsilon_2=const$. 

As it follows from the above discussion, both of the equations at 
a constant value of $q_{z,j}$ have the solution in the form of Eq. (4). 
However, compared to the case of the resonant Wood's anomaly, where 
$k_z=const$, the surface plasmon-polariton resonance is characterized 
with strong frequency dependence of $q_{z,j}$. In the latter case 
the spectral angular dependence is obtained by solving the equation 
\begin{equation}
{k_0}^2(\varepsilon_s-{\sin}^{2}\theta)-2k_{0}G\sin{\theta}
\cos{\varphi}-G^2=0\:,
\end{equation}
which reduces to that for a fourth-order polynomial function of $k_0$
from which one gets $\lambda_{SP}=2\pi/k_0$. From Eq. (8) one can also 
derive the dispersion relation $k_0(k_{||})$ for the surface 
plasmon-polaritons:
\begin{eqnarray}
&&{k_0}^2 = p/2+g(k_{||})(\varepsilon_2+1)/(2\varepsilon_2)\pm \\&&
\sqrt{(p/2)^2+g(k_{||})^2(\varepsilon_2+1)^2/(2\varepsilon_2)^2+
pg(k_{||})(\varepsilon_2-1)/(2\varepsilon_2)}\:. \nonumber
\end{eqnarray}
The analysis of the dispersion relations made by means of Eqs.
(7) and (9) enables us to distinguish between RWA and SPPR. 
Though both expressions give almost straight lines $k_0(k_{||})$,
their slope and position can be quite different (the latter can be seen
from the corresponding values of $k_0$ at $\theta=0$).

\section {Discussion (connection to the scattering matrix)}

The theory presented in this paper is based on Eq.~(1), which
is applied to dielectric and plasmonic grating structures. Therefore 
it may be useful to provide some additional explanation on this point.
Let us consider the scattering matrix $S$ (2$\times$2) for a single
plane layer with the refractive index $n=\sqrt{\tilde{\varepsilon}}$,
which for simplicity can be taken real (in the absence of absorption)
[\onlinecite{Tikhodeev2002}]. The elements of the matrix $S$ can be
written in terms of the amplitude reflection and transmission
coefficients, $r$ and $t$, for an incident plane wave. It is well-known
that the equation $det(S^{-1})=(r^2-t^2)^{-1}=0$ determines the
eigenfrequencies $\Omega=\Omega'+i\Omega''$ and eigenvalues $k_z=k_z'+ik_z''$ 
of the $z$-component of the wave vector inside the layer. Hence, using the 
expressions for $r$ and $t$, one can show that $1-r_{21}r_{23}e^{2ik_zd}=0$,
where $r_{21}$ and $r_{23}$ are the coefficients of reflection from
two interfaces and $d$ is the layer thickness. In the case of a guided 
mode the light wave experiences total internal reflection from the 
interfaces, so that $|r_{21}|=|r_{23}|=1$ and, consequently, $k_z$ is real.
The magnitude $k$ of the wave vector of light propagating in the layer
satisfies the equation 
\begin{equation}
k^2=(\omega n/c)^2=k_{||}^2+k_z^2\:.
\end{equation}
Due to the conservation of the tangential component $k_{||}$ it can be 
represented as $k_{||}=k_0\sin\theta_0$, where $k_0=\omega/c$ is the
magnitude of the wave vector in vacuum, when a plane monochromatic wave
of the frequency $\omega$ falls onto the layer–vacuum interface from vacuum
at an angle $\theta_0$ (and reflects at the angle $\theta=\theta_0$);  since 
$k_z''=0$, from Eq. (10) one gets $\omega=\Omega'$. Because of the 
periodicity of the structure the wave corresponding to a guided mode is 
scattered and goes outside the layer. This is taken into account by the
replacement of ${\bf k_{||}}$ in Eq. (10) by ${\bf k_{||}}+{\bf G}$, 
that eventually leads to Eq. (1).  

\section {Conclusion}
In conclusion, we theoretically studied some peculiarities of the
diffraction condition for the resonant Wood's anomaly obtained in the
approximation of a single scattering vector and by using the effective
parameters of a quasi-guided mode. The expression for location of the
spectral peaks in the reflection and transmission spectra of plasmonic
grating structures is given; if the dispersion of the transverse component
of the wave vector in the waveguide layer can be ignored, this expression 
is converted to the analogous one for dielectric grating structures as
the plasma frequency tends to zero. This approach can be useful in 
analyzing spectral angular dependencies obtained from the reflection
and transmission spectra demonstrating the resonant Wood's anomaly and
surface plasmon-polariton resonance. In practice, the difference between
these two types of resonances can be established by simulating the 
experimental data by numerical calculation with a subsequent model
calculation performed for the same parameters by neglecting the 
absorption in the grating structure and then by fitting the obtained
points to Eqs. (6-9).

\begin{acknowledgments}
The author is grateful to S. A. Dyakov and A. B. Pevtsov for helpful 
discussions. 
\end{acknowledgments}

\end{document}